# Giant perpendicular magnetic anisotropy enhancement in MgO-based magnetic tunnel junction by using Co/Fe composite layer

Libor Vojáček[a*†], Fatima Ibrahim[a], Ali Hallal[a], Bernard Dieny[a], Mairbek Chshiev[a,b]*

[a]Univ. Grenoble Alpes, CNRS, CEA, Spintec, 38000 Grenoble, France

[b]Institut Universitaire de France (IUF)

KEYWORDS: perpendicular magnetic tunnel junctions, perpendicular magnetic anisotropy enhancement, strain-induced anisotropy, STT-MRAM

ABSTRACT: Magnetic tunnel junctions with perpendicular anisotropy form the basis of the spin-transfer torque magnetic random-access memory (STT-MRAM), which is non-volatile, fast, dense, and has quasi-infinite write endurance and low power consumption. Based on density functional theory (DFT) calculations, we propose an alternative design of magnetic tunnel junctions comprising Fe(n)Co(m)Fe(n)|MgO storage layers with greatly enhanced perpendicular magnetic anisotropy (PMA) up to several mJ/m$^2$, leveraging the interfacial perpendicular anisotropy of Fe|MgO along with a stress-induced bulk PMA discovered within bcc Co. This giant enhancement dominates the demagnetizing energy when increasing the film thickness. The tunneling magnetoresistance (TMR) estimated from the Julliere model is comparable with that of the pure Fe|MgO case. We discuss the advantages and pitfalls of a real-life fabrication of the structure and propose the Fe(3ML)Co(4ML)Fe(3ML) as a storage layer

for MgO-based STT-MRAM cells. The large PMA in strained bcc Co is explained in the framework of Bruno's model by the MgO-imposed strain and consequent changes in the energies of $d_{yz}$ and $d_{z2}$ minority-spin bands.

## 1. Introduction

MgO-based magnetic tunnel junctions (MTJs) are used in today's hard-disk drive read heads and a variety of magnetic field sensors for their supremely high tunneling magnetoresistance (TMR) effect[1]. Hard-disk drives, however, are approaching their scaling limits[2]. Besides, spin-transfer torque magnetic random-access memory (STT-MRAM), also based on MTJs comprising an MgO tunnel barrier, is entering into volume production for eFLASH replacement and SRAM type of applications. They are non-volatile, fast (5-50ns cycle time), can be made relatively dense (Gbit), with low power consumption (100fJ/write event), and exhibit very good write endurance[3–5].

The building block of STT-MRAM is a cell with (1) high TMR for good readability, (2) high spin-transfer torque efficiency for good writability, and (3) high magnetic anisotropy for good thermal stability and therefore memory retention[6,7] . All of these requirements must be satisfied together which is the case in perpendicularly magnetized CoFeB|MgO MTJs as long as the cell diameter remains larger than ~30nm[8]. Below this diameter, the perpendicular anisotropy provided by CoFeB|MgO interface becomes too weak in regards to thermal fluctuations so that the memory retention reduces excessively. In this work, we focus on improving the third requirement – the perpendicular magnetic anisotropy (PMA) – of the MgO-based MTJ, therefore allowing improved downsize scalability of out-of-plane magnetized MRAM.

Although heavy metals like Pt or Pd can enhance PMA[9,10], they do so by increasing the spin-orbit coupling (SOC) parameter ξ. This is, however, associated with the undesirable side effect of increasing the Gilbert damping[11], thus increasing the spin-transfer torque switching current[12,13]. To avoid this problem, recipes based on purely 3d metallic elements were developed[14,15]. However, these recipes are based on Fe-Ni or Co-Ni alternating atomic layers, yielding structures that are intrinsically complex to fabricate or may require excessively high deposition/annealing temperature. They can also get deteriorated upon the annealing required to obtain good crystallization of the MgO barrier and surrounding magnetic electrodes.

In this work, we propose a different approach based on introducing a bulk Co interlayer into a simple Fe|MgO MTJ. The latter exhibits comparable or stronger PMA than the aforementioned Fe-Ni or Co-Ni alternating atomic layers. In addition, the PMA characterized by the magnetocrystalline anisotropy energy $E_{\text{MCA}}$ increases with the film thickness at comparable or higher rate. Lastly, the Co Curie temperature (1404K) is significantly higher than that of Fe (1043K) and twice higher than that of Ni (631K), which provides higher temperature stability[16].

The paper is organized as follows. In Section 2.1, based on DFT calculations, we propose a new type of MTJ with an enhanced PMA and high TMR. In Section 2.2, we discuss the real-life fabrication aspect. In Section 2.3, the systematic calculations supporting our proposal are presented. In Section 2.4, the large bulk Co|MgO PMA is explained by the electronic structure and Bruno's model.

## 2. Results and Discussion

### 2.1. MTJ with greatly enhanced PMA

The DFT calculations were performed using the Vienna ab initio simulation package (VASP)[17,18]. Besides the electronic structure, the basic output of the calculation is the magnetocrystalline anisotropy energy $E_{\text{MCA}}$ and its atomic site-resolved contributions. Positive (negative) value of $E_{\text{MCA}}$ indicates PMA (in-plane anisotropy), respectively. Calculation details are given in the Supporting Information.

In this study, we find that a significant contribution to PMA originates from the bulk of epitaxial bcc Co on top of MgO. Its origin will be presented in detail in Sections 2.3 and 2.4. We exploit this finding and propose to enhance the PMA of conventional Fe|MgO MTJ by replacing the bulk Fe layers with Co. The effect can be further enhanced by sandwiching the magnetic layer between two MgO barriers. The proposed improved MTJ storage layer thus takes the form MgO|Fe(n)Co(m)Fe(n)|MgO with $n \geq 2$ and $m \geq 3$, as shown in Fig. 1(c). It is required to have at least two Fe atoms at the MgO interface and three successive Co atoms in the bulk to obtain the PMA enhancement (for details see Fig. S1 in the Supporting Information).

Structures with different n and m were systematically investigated. The thickness of MgO in all the calculations was chosen to be 5 (6) monolayers for odd (even) number of metal layers, respectively. Fig. 1(a) shows the effective PMA comprising the magnetocrystalline anisotropy energy $E_{MCA}$ and dipole-dipole induced demagnetization energy $E_{dd}$ as a function of n, m. One can clearly see that the effective PMA does not vanish with increasing thickness but interestingly it grows steadily. This is in contrast to the pure Fe|MgO case (grey line), where the demagnetizing energy $E_{dd}$ drives the magnetization in-plane for thicknesses above 11 monolayers (ML). The variation of $E_{MCA}$ and $E_{dd}$ as a function of m for n=2 is shown in Fig. 1(b).

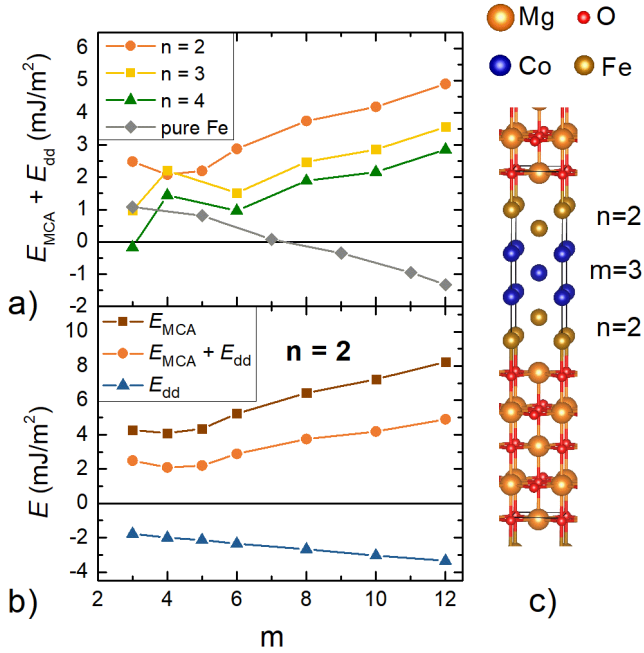

**Figure 1.** (a) Effective PMA in MgO|Fe(n)Co(m)Fe(n)|MgO as a function of n, m. Pure MgO|Fe|MgO with a thickness of m+4 ML is shown by gray symbols for comparison (the same as the overall thickness for n = 2). (b) $E_{MCA}$, $E_{dd}$ and their sum for n = 2. (c) Supercell of the MgO|Fe2Co3Fe2|MgO with periodic boundary conditions applied in all directions (produced by VESTA[19]).

The dipole-dipole energy $E_{dd}$ is obtained by summing over all the dipole-dipole interactions[20,21] (see the Supporting Information for details). $E_{dd}$ acts mainly as a demagnetizing energy[20], favoring the in-plane magnetization orientation. $E_{dd}$ is proportional to the product of the magnetic moments of the interacting atoms ($\mu_1 \cdot \mu_2$). Thus, replacing the bulk Fe with Co decreases its magnitude knowing that in the bulk of the layer $\mu_{Fe} \approx 2.5~\mu_B$ while $\mu_{Co} \approx 1.8~\mu_B$, where $\mu_B$ is the Bohr magneton. Thus, the effect of the bulk Co is two-fold: it increases the positive $E_{MCA}$ (discussed in detail in Sections 2.3 and 2.4) and diminishes the negative demagnetizing dipole-dipole contribution.

Replacing the MgO on one side of the metal film with vacuum, as tested on the n=4, m=6 structure, decreases the $E_{MCA}$ by 16% (see Fig. S2 in the Supporting Information). This is predominantly due to the drop in $E_{MCA}$ of the two interfacial Fe atoms[23]. The effect of replacing the MgO by vacuum in pure bcc Fe(Co,Ni)|MgO is shown in Fig. S3 in the Supporting Information. However, when the metal thicknesses are reduced (n=2, m=3), this drop is 36% and reaches even 86% when both interfaces are in contact with vacuum instead of MgO. This explains why Hotta et al.[14] observed no enhancement of anisotropy in Fe2Co3Fe2. They mimicked the presence of MgO by setting the lattice parameter equal to that of bulk MgO while in reality having vacuum at the interfaces. Additionally, changing the MgO thickness from 3 to 7ML has a negligible effect on the $E_{MCA}$ in the order of 0.01 mJ/m$^2$, as tested on the Fe3Co3Fe2.

Since we are interested in implementing this proposed storage layer in a full MTJ stack, we investigated its expected TMR amplitude. A large TMR of 410% at room temperature has been observed previously in pure bcc Co|MgO|Co MTJs[22]. In addition, Co in combination with Fe is often used for its record-holding TMR values. Therefore, we expect the high TMR to be present also in the proposed Fe|MgO MTJs with the inserted Co bulk layer. The TMR value was estimated from the Julliere formula[23] with the spin polarization calculated from the local density of states at the Fermi level $E_F$ of the interfacial Fe atoms. The values are very similar to those of pure Fe|MgO, i.e. around 400%. Because the PMA does not originate solely from the interface in the proposed structure, in contrast to Fe|MgO[24], this suggests that we could separately tune the PMA by the bulk Co and the TMR by the Fe|MgO interface.

## 2.2. Towards practical implementation

As mentioned in the introduction, the proposed MTJ is conceptually much simpler and robust against annealing than the alternating layers-based MTJ[14,15] with similar properties. However, there are two main issues that we address regarding the fabrication of our structure, namely the stability of the bcc Co phase and the robustness against the Fe-Co interface not being atomically sharp.

Although the natural form for Co is hcp, the metastable bcc Co phase can be grown at room temperature[25–27]. It has been successfully grown on top of Fe with thickness up to 15 ML[28], with well-defined interfaces and no visible interdiffusion. The observed strain of 10% in bcc Co|MgO is very large, but still in the limit of what is experimentally realizable[29]. Indeed, Yuasa *et al.*[22] fabricated bcc Co(4ML)|MgO(10ML)|Co(4ML) MTJ and measured a record-holding TMR of 410% at room temperature. The bcc phase stability will probably be highest if the device is used as a double-barrier MTJ, which also provides the highest PMA from the interfacial Fe. This is based on our structural relaxation showing that the bcc Co is preserved on top of MgO, while it transforms into the fcc phase when surrounded by vacuum.

Sharpness of the Fe-Co interface is another important factor to consider. From the simulations, it follows that any interdiffusion is fatal for the PMA when the Fe or Co thickness is less than 2ML or 3ML, respectively (i.e $n < 2$, $m < 3$; see Fig. S1 in the Supporting information). Robustness will clearly be achieved at larger Fe and Co thicknesses. Larger Co thickness is favorable as it increases the PMA [Fig. 1], but thicker bcc Co will probably be harder to fabricate[22]. Higher Fe thickness also adds the desired robustness, but the PMA is decreased as shown in Fig. 1(a) [for layer-resolved behavior see Fig. S4 in the Supporting Information]. In addition, the stability of the bcc Co phase might be increased with thicker Fe, as it is generally easier to grow bcc Co on Fe than on MgO. Besides, Co does not wet well on oxides due to its high surface tension while Fe does[30,31]. Looking at Fig. 1(a) and considering all the mentioned aspects, the MgO|Fe(3ML)Co(4ML)Fe(3ML)|MgO seems a promising candidate as a storage layer for STT-MRAM cells with much improved thermal stability compared to conventional STT-MRAM.

Indeed, when the storage layer is sandwiched between two MgO layers, the anisotropy per unit area is of the order of 2 mJ/m² from the interfacial contribution minus ~1.2 mJ/m² from demagnetizing energy (dependent on the chosen storage layer thickness) yielding a net effective anisotropy per unit area ~0.8mJ/m².[4] In comparison, the net anisotropy per unit area in the proposed structure is ~2.2mJ/m² being almost 3 times larger. This means that for the same thermal stability factor, the cell area could be reduced by a factor of 3 in the proposed structure compared to conventional MRAM[7,32].

## 2.3. PMA in bcc Fe(Co, Ni)|MgO thin films

The idea of the improved MTJ proposed above was driven by our systematic investigation of the thickness dependence of $E_{MCA}$ in pure bcc (001) Fe, Co and Ni |MgO ultrathin films. The $E_{MCA}$ as a function of metallic layer thickness is presented in Fig. 2(a). While for Fe the $E_{MCA}$ converges to a constant value[24], we observe a steady increase for Co. The behavior for Ni is more subtle. To elucidate why the trend varies among the three metals, we show in Fig. 2(b) the layer-resolved contributions to the $E_{MCA}$ (i.e. the contributions from each atomic layer separately).

For Fe, the main contribution to $E_{MCA}$ comes from the first two interfacial layers[24,33]. Increasing the thickness does not affect the electronic properties of the interfacial layers in a significant way[24] (see Fig. S5 in the Supporting Information). The bulk layers almost do not contribute to the PMA, hence the $E_{MCA}$ does not change.

On the contrary, all the bulk layers of Co seem to contribute with a significant positive $E_{MCA}$ value as evident from Fig. 2(b). Hence, the $E_{MCA}$ grows almost linearly with the number of added bulk Co layers. This observation is the cornerstone of this paper.

For Ni, the influence of the interface manifests itself as deep as 6 monolayers, with the two interfacial monolayers contributing a negative $E_{MCA}$. This is the reason for the in-plane preference in the 5ML structure as shown in Fig. 2(a). Although the deeper bulk layers contribute positively, the $E_{MCA}$ does not grow monotonically as expected because the interfacial contributions change upon thickness increase (see Fig. S5 in the Supporting Information). The bcc Ni |MgO is also problematic because of the large strain of ≈ 15%.

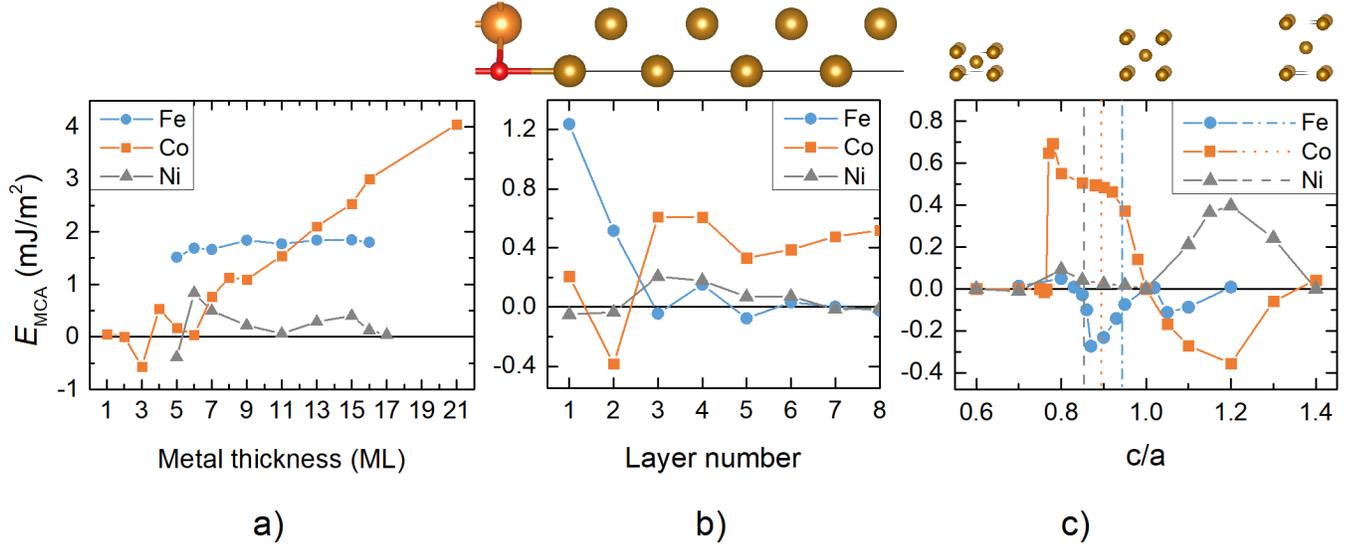

**Figure 2.** (a) Thickness dependence of $E_{MCA}$ in bcc Fe(Co, Ni)|MgO. Note that the PMA in Co increases steadily. (b) Layer-resolved $E_{MCA}$ in the structure with 15ML of metal. Layer 1 is the interfacial layer, the layer number increases towards the bulk of the material. The largest contribution for Fe comes from the interface, for Co it comes from the bulk. (c) $E_{MCA}$ in purely bulk bcc Fe, Co and Ni as a function of c/a ratio. Dashed lines indicate the typical value of c/a in the bulk of the given metal|MgO (see text for details). At its typical strain, the $E_{MCA}$ for Co is the same as in the bulk layers in Fig. 2.

### 2.4. Strain-induced anisotropy and Bruno's model

The large positive bulk $E_{MCA}$ in Co|MgO is caused by the strain that is induced within the Co by the MgO. To confirm this hypothesis, we have calculated $E_{MCA}$ as a function of c/a ratio in the primitive bcc unit cell for each of the metals shown in Fig. 2(c). The a and b lattice parameters were set to the value of the relaxed bulk bcc Fe, Co or Ni unit cell, while the c lattice parameter was varied. The typical c/a ratios we found within the bulk metal layers in Fe, Co and Ni interfaced with MgO are 0.94, 0.89, and 0.85, respectively [dashed lines in Fig. 2(c)].

It clearly follows from Fig. 2(c) that for Fe and Ni, this bulk strain-induced $E_{MCA}$ is very small at their typical bulk strains (dashed lines). However, we observe a large positive contribution of $\approx 0.5$ mJ/m² for Co, which is the same value expected from Fig. 2(b). Indeed, artificially setting c/a = 1 in a previously relaxed Co|MgO eliminates the bulk contribution to PMA, thus confirming that strain plays the central role. We note that the calculated $E_{MCA}$(c/a) dependence in Fig. 2(c) corresponds well to previous findings[34,35], where the focus was limited to c/a > 1.

Deeper analysis shows that the c/a dependence of $E_{MCA}$ for c/a < 1 in bcc Co|MgO can be well explained in the framework of Bruno's model[36,37]. Bruno's model treats magnetic anisotropy based on the second-order perturbation theory, so it involves interactions of filled and unfilled orbitals, separated by the Fermi level. Depending on the two interacting $d$ orbitals $\mu$ and $\mu$', there are 10 nonzero contributions to $E_{MCA}$ that are either positive or negative dictated by the respective $P_{\mu\mu'}$ matrix component (see the Supporting information for details on $P_{\mu\mu'}$ calculation). Furthermore, the strength of the contribution is inversely proportional to the difference between the energies of the two interacting orbitals. Because the majority-spin band is almost fully occupied, only the minority-spin states are taken into account.[37]

Let us now reproduce the Co $E_{MCA}$ dependence as a function of c/a from Fig. 2(c) by applying Bruno's model to the band structure at high-symmetry Z point using $\xi_{Co}$ = 84 meV for the spin-orbit coupling parameter[38]. The resulting 'Bruno' curve is shown in Fig. 3(a). For comparison, we plot the Co $E_{MCA}$ curve ('DFT') from Fig. 2(c). In the region of interest, around c/a = 0.90, these two show good correspondence.

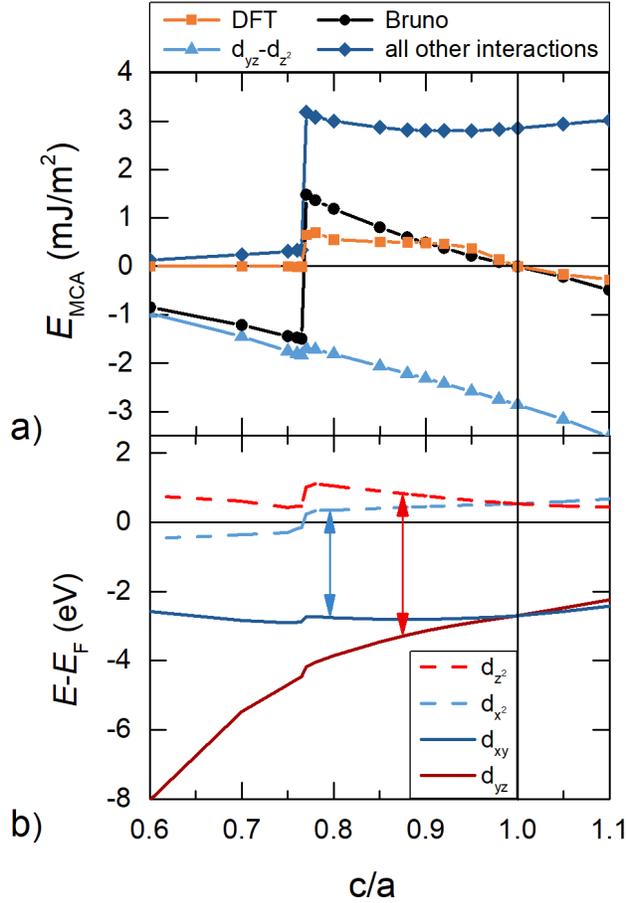

**Figure 3.** (a) The Co $E_{MCA}$ dependence on c/a obtained by ('DFT') from Fig. 2(c) compared with the one calculated with Bruno's model at the high-symmetry Z point ('Bruno'). They correspond well around c/a = 0.90. The '$d_{yz}$-$d_{z^2}$' interaction contribution to 'Bruno' and the sum of 'all (the nine) other contributions' are separately shown. The $d_{yz}$-$d_{z^2}$ clearly dictates the changes in $E_{MCA}$ while the sum of all the other interactions varies only little with c/a. (b) The bcc Co c/a dependent band-structure at the high-symmetry

Z point around $E_F$. The energy difference between $d_{z2}$ and $d_{yz}$ increases with decreasing c/a, resulting in diminishing this negative '$d_{yz}$-$d_{z2}$' contribution in Fig. 3(a).

Inspecting the individual nonzero contributions to the 'Bruno' curve, we notice that the $d_{yz}$-$d_{z2}$ is the dominant one. Indeed, in Fig. 3(a) the '$d_{yz}$-$d_{z2}$' curve reflects the c/a dependence of the 'Bruno' curve, while the sum of all other interactions stays almost constant around c/a = 0.90. Note that the interaction of the filled $d_{yz}$ band with the unfilled $d_{z2}$ band has the largest negative $P_{\mu\mu'}$ weight coefficient (see the Supporting Information). To elucidate the $d_{yz}$-$d_{z2}$ trend, in Fig. 3(b) we show the band structure around Fermi energy at the high-symmetry Z point. We observe that the energy difference between the $d_{yz}$ and $d_{z2}$ bands increases for c/a < 1. Because the contribution to $E_{MCA}$ is inversely proportional to the energy difference of the interacting orbitals, this negative value diminishes with decreasing c/a, which increases the overall $E_{MCA}$. In conclusion, the large bulk Co PMA is caused by the increase of the energy difference between the filled $d_{yz}$ and unfilled $d_{z2}$ minority-spin bands, which in turn is caused by the MgO-imposed strain. While we illustrate this on the high-symmetry Z point, the c/a trend is maintained when all the k-points in the Brillouin zone are integrated over.

## 3. Conclusion

We proposed an alternative concept of MTJ with strongly enhanced perpendicular magnetic anisotropy based on introducing a Co interlayer into the bulk of conventional Fe|MgO MTJ. DFT calculations confirm that the PMA enhancement overcomes the negative demagnetizing energy in these Fe(n)Co(m)Fe(n)|MgO structures. The TMR shows values similar to the pure Fe|MgO case. There is a trade-off between the enhancement magnitude and its robustness against the Fe-Co interfacial diffusion

in a prospective real-life fabrication process. The Fe(3 ML)Co(4 ML)Fe(3 ML) seems of strong potential as a storage layer for MgO-based STT-MRAM cells. This design is based on the presented systematic study of PMA in bcc Fe(Co,Ni)|MgO, showing clearly that the MgO-imposed compressive strain induces a significant bulk PMA in bcc Co. This phenomenon is explained in the framework of Bruno's model and attributed to strain-induced changes in the energies of the minority-spin filled $d_{yz}$ and unfilled $d_{z2}$ orbitals at the Fermi level.

ASSOCIATED CONTENT

Details on the DFT and demagnetization energy calculation methods, layer-resolved $E_{MCA}$ for structures not complying with the n ≥ 2, m ≥ 3 requirement, layer-resolved $E_{MCA}$ for Fe(n)Co(m)Fe(n) structures with vacuum at the interface, for pure bcc 16ML Fe(Co,Ni)|(MgO/vacuum), for Fe(n)Co(m)Fe(n) with various n and m, and finally for pure bcc Fe(Co,Ni)|MgO as a function of thickness. (PDF)

Crystallographic structure of the relaxed Fe2Co3Fe2|MgO (CIF)

AUTHOR INFORMATION

**Corresponding Authors**

* Libor Vojáček – Univ. Grenoble Alpes, CNRS, CEA, Spintec, 38000 Grenoble, France; Email: Libor.Vojacek@vutbr.cz

* Mairbek Chshiev – Univ. Grenoble Alpes, CNRS, CEA, Spintec, 38000 Grenoble, France; Email: mair.chshiev@cea.fr

**Authors**

Fatima Ibrahim – Univ. Grenoble Alpes, CNRS, CEA, Spintec, 38000 Grenoble, France

Ali Hallal – Univ. Grenoble Alpes, CNRS, CEA, Spintec, 38000 Grenoble, France

Bernard Dieny – Univ. Grenoble Alpes, CNRS, CEA, Spintec, 38000 Grenoble, France

**Present Addresses**

†Libor Vojáček – CEITEC BUT, Brno University of Technology, Purkyňova 123, 612 00 Brno, Czech Republic

**Funding Sources**

Co-funded by the Erasmus+ Programme of the European Union.

**Notes**

The authors declare no competing financial interest.

ACKNOWLEDGMENT

Computational resources were partially supplied by the project "e-Infrastruktura CZ" (e-INFRA LM2018140) provided within the program Projects of Large Research, Development and Innovations Infrastructures. BD acknowledges ERC funding via ERC Adv grant MAGICAL 669204.

ABBREVIATIONS

DFT density functional theory; MTJ magnetic tunnel junction; PMA perpendicular magnetic anisotropy; TMR tunneling magnetoresistance; ML monolayer; SOC spin-orbit coupling; STT-MRAM spin-transfer torque magnetic random-access memory.

REFERENCES

(1) Bajorek, C. H. Magnetoresistive (MR) Heads and the Earliest MR Head-Based Disk Drives: Sawmill and Corsair. *Comput. Hist. Mus. Mt. View CA Tech Rep* **2014**, 11.

(2) Richter, H. J. Density Limits Imposed by the Microstructure of Magnetic Recording Media. *J. Magn. Magn. Mater.* **2009**, *321* (6), 467–476. https://doi.org/10.1016/j.jmmm.2008.04.161.

(3) Meena, J.; Sze, S.; Chand, U.; Tseng, T.-Y. Overview of Emerging Nonvolatile Memory Technologies. *Nanoscale Res. Lett.* **2014**, *9* (1), 526. https://doi.org/10.1186/1556-276X-9-526.

(4) Khvalkovskiy, A. V.; Apalkov, D.; Watts, S.; Chepulskii, R.; Beach, R. S.; Ong, A.; Tang, X.; Driskill-Smith, A.; Butler, W. H.; Visscher, P. B.; Lottis, D.; Chen, E.; Nikitin, V.; Krounbi, M. Basic Principles of STT-MRAM Cell Operation in Memory Arrays. *J. Phys. Appl. Phys.* **2013**, *46* (7), 074001. https://doi.org/10.1088/0022-3727/46/7/074001.

(5) Huai, Y. Spin-Transfer Torque MRAM (STT-MRAM): Challenges and Prospects. *AAPPS Bull.* **2008**, *18* (6), 8.

(6) Kent, A. D. Perpendicular All the Way. *Nat. Mater.* **2010**, *9* (9), 699–700. https://doi.org/10.1038/nmat2844.

(7) Dieny, B.; Chshiev, M. Perpendicular Magnetic Anisotropy at Transition Metal/Oxide Interfaces and Applications. *Rev. Mod. Phys.* **2017**, *89* (2), 025008. https://doi.org/10.1103/RevModPhys.89.025008.

(8) Ikeda, S.; Miura, K.; Yamamoto, H.; Mizunuma, K.; Gan, H. D.; Endo, M.; Kanai, S.; Hayakawa, J.; Matsukura, F.; Ohno, H. A Perpendicular-Anisotropy CoFeB–MgO Magnetic Tunnel Junction. *Nat. Mater.* **2010**, *9* (9), 721–724. https://doi.org/10.1038/nmat2804.

(9) Carcia, P. F.; Meinhaldt, A. D.; Suna, A. Perpendicular Magnetic Anisotropy in Pd/Co Thin Film Layered Structures. *Appl. Phys. Lett.* **1985**, *47* (2), 178–180. https://doi.org/10.1063/1.96254.

(10) Hashimoto, S.; Ochiai, Y.; Aso, K. Perpendicular Magnetic Anisotropy and Magnetostriction of Sputtered Co/Pd and Co/Pt Multilayered Films. *J. Appl. Phys.* **1989**, *66* (10), 4909–4916. https://doi.org/10.1063/1.343760.

(11) Bruno, P. Physical Origins and Theoretical Models of Magnetic Anisotropy. In *Ferienkurse des Forschungszentrums Julich*; 1993; p 29.

(12) Mizukami, S.; Sajitha, E. P.; Watanabe, D.; Wu, F.; Miyazaki, T.; Naganuma, H.; Oogane, M.; Ando, Y. Gilbert Damping in Perpendicularly Magnetized Pt/Co/Pt Films Investigated by All-Optical Pump-Probe Technique. *Appl. Phys. Lett.* **2010**, *96* (15), 152502. https://doi.org/10.1063/1.3396983.

(13) Slonczewski, J. C. Current-Driven Excitation of Magnetic Multilayers. *J. Magn. Magn. Mater.* **1996**, *159* (1–2), L1–L7. https://doi.org/10.1016/0304-8853(96)00062-5.

(14) Hotta, K.; Nakamura, K.; Akiyama, T.; Ito, T.; Oguchi, T.; Freeman, A. J. Atomic-Layer Alignment Tuning for Giant Perpendicular Magnetocrystalline Anisotropy of 3 d Transition-Metal Thin Films. *Phys. Rev. Lett.* **2013**, *110* (26), 267206. https://doi.org/10.1103/PhysRevLett.110.267206.

(15) Nakamura, K.; Ikeura, Y.; Akiyama, T.; Ito, T. Giant Perpendicular Magnetocrystalline Anisotropy of 3 *d* Transition-Metal Thin Films on MgO. *J. Appl. Phys.* **2015**, *117* (17), 17C731. https://doi.org/10.1063/1.4916191.

(16) Cullity, B. D. *Introduction to Magnetic Materials*; Addison-Wesley series in metallurgy and materials; Addison-Wesley: Reading, Mass., 1972.

(17) Kresse, G.; Hafner, J. *Ab Initio* Molecular Dynamics for Liquid Metals. *Phys. Rev. B* **1993**, *47* (1), 558–561. https://doi.org/10.1103/PhysRevB.47.558.

(18) Kresse, G.; Furthmüller, J. Efficiency of Ab-Initio Total Energy Calculations for Metals and Semiconductors Using a Plane-Wave Basis Set. *Comput. Mater. Sci.* **1996**, *6* (1), 15–50. https://doi.org/10.1016/0927-0256(96)00008-0.

(19) Momma, K.; Izumi, F. *VESTA 3* for Three-Dimensional Visualization of Crystal, Volumetric and Morphology Data. *J. Appl. Crystallogr.* **2011**, *44* (6), 1272–1276. https://doi.org/10.1107/S0021889811038970.

(20) Draaisma, H. J. G.; de Jonge, W. J. M. Surface and Volume Anisotropy from Dipole-dipole Interactions in Ultrathin Ferromagnetic Films. *J. Appl. Phys.* **1988**, *64* (7), 3610–3613. https://doi.org/10.1063/1.341397.

(21) Daalderop, G. H. O.; Kelly, P. J.; Schuurmans, M. F. H. First-Principles Calculation of the Magnetocrystalline Anisotropy Energy of Iron, Cobalt, and Nickel. *Phys. Rev. B* **1990**, *41* (17), 11919–11937. https://doi.org/10.1103/PhysRevB.41.11919.

(22) Yuasa, S.; Fukushima, A.; Kubota, H.; Suzuki, Y.; Ando, K. Giant Tunneling Magnetoresistance up to 410% at Room Temperature in Fully Epitaxial Co∕MgO∕Co Magnetic Tunnel Junctions with Bcc Co(001) Electrodes. *Appl. Phys. Lett.* **2006**, *89* (4), 042505. https://doi.org/10.1063/1.2236268.

(23) Julliere, M. Tunneling between Ferromagnetic Films. *Phys. Lett. A* **1975**, *54* (3), 225–226. https://doi.org/10.1016/0375-9601(75)90174-7.

(24) Hallal, A.; Yang, H. X.; Dieny, B.; Chshiev, M. Anatomy of Perpendicular Magnetic Anisotropy in Fe/MgO Magnetic Tunnel Junctions: First-Principles Insight. *Phys. Rev. B* **2013**, *88* (18), 184423. https://doi.org/10.1103/PhysRevB.88.184423.

(25) Li, H.; Tonner, B. P. Direct Experimental Identification of the Structure of Ultrathin Films of Bcc Iron and Metastable Bcc and Fcc Cobalt. *Phys. Rev. B* **1989**, *40* (15), 10241–10248. https://doi.org/10.1103/PhysRevB.40.10241.

(26) Subramanian, S.; Liu, X.; Stamps, R. L.; Sooryakumar, R.; Prinz, G. A. Magnetic Anisotropies in Body-Centered-Cubic Cobalt Films. *Phys. Rev. B* **1995**, *52* (14), 10194–10201. https://doi.org/10.1103/PhysRevB.52.10194.

(27) Liu, X.; Stamps, R. L.; Sooryakumar, R.; Prinz, G. A. Magnetic Anisotropies in Thick Body Centered Cubic Co. *J. Appl. Phys.* **1996**, *79* (8), 5387. https://doi.org/10.1063/1.362312.

(28) Houdy, Ph.; Boher, P.; Giron, F.; Pierre, F.; Chappert, C.; Beauvillain, P.; Dang, K. L.; Veillet, P.; Velu, E. Magnetic and Structural Properties of Rf-sputtered Co/Fe and Co/Cr Multilayers. *J. Appl. Phys.* **1991**, *69* (8), 5667–5669. https://doi.org/10.1063/1.347930.

(29) Sander, D. The Magnetic Anisotropy and Spin Reorientation of Nanostructures and Nanoscale Films. *J. Phys. Condens. Matter* **2004**, *16* (20), R603–R636. https://doi.org/10.1088/0953-8984/16/20/R01.

(30) Dieny, B.; Sankar, S.; McCartney, M. R.; Smith, D. J.; Bayle-Guillemaud, P.; Berkowitz, A. E. Spin-Dependent Tunneling in Discontinuous Metal/Insulator Multilayers. *J. Magn. Magn. Mater.* **1998**, *185* (3), 283–292. https://doi.org/10.1016/S0304-8853(98)00028-6.

(31) Fahsold, G.; Pucci, A.; Rieder, K.-H. Growth of Fe on MgO(001) Studied by He-Atom Scattering. *Phys. Rev. B* **2000**, *61* (12), 8475–8483. https://doi.org/10.1103/PhysRevB.61.8475.

(32) Apalkov, D.; Dieny, B.; Slaughter, J. M. Magnetoresistive Random Access Memory. *Proc. IEEE* **2016**, *104* (10), 1796–1830. https://doi.org/10.1109/JPROC.2016.2590142.

(33) Yang, H. X.; Chshiev, M.; Dieny, B.; Lee, J. H.; Manchon, A.; Shin, K. H. First-Principles Investigation of the Very Large Perpendicular Magnetic Anisotropy at Fe | MgO and Co | MgO Interfaces. *Phys. Rev. B* **2011**, *84* (5), 054401. https://doi.org/10.1103/PhysRevB.84.054401.

(34) Burkert, T.; Eriksson, O.; James, P.; Simak, S. I.; Johansson, B.; Nordström, L. Calculation of Uniaxial Magnetic Anisotropy Energy of Tetragonal and Trigonal Fe, Co, and Ni. *Phys. Rev. B* **2004**, *69* (10), 104426. https://doi.org/10.1103/PhysRevB.69.104426.

(35) Burkert, T.; Nordström, L.; Eriksson, O.; Heinonen, O. Giant Magnetic Anisotropy in Tetragonal FeCo Alloys. *Phys. Rev. Lett.* **2004**, *93* (2), 027203. https://doi.org/10.1103/PhysRevLett.93.027203.

(36) Bruno, P. Tight-Binding Approach to the Orbital Magnetic Moment and Magnetocrystalline Anisotropy of Transition-Metal Monolayers. *Phys. Rev. B* **1989**, *39* (1), 865–868. https://doi.org/10.1103/PhysRevB.39.865.

(37) Zhang, J.; Lukashev, P. V.; Jaswal, S. S.; Tsymbal, E. Y. Model of Orbital Populations for Voltage-Controlled Magnetic Anisotropy in Transition-Metal Thin Films. *Phys. Rev. B* **2017**, *96* (1), 014435. https://doi.org/10.1103/PhysRevB.96.014435.

(38) Popescu, V.; Ebert, H.; Nonas, B.; Dederichs, P. H. Spin and Orbital Magnetic Moments of 3 d and 4 d Impurities in and on the (001) Surface of Bcc Fe. *Phys. Rev. B* **2001**, *64* (18), 184407. https://doi.org/10.1103/PhysRevB.64.184407.

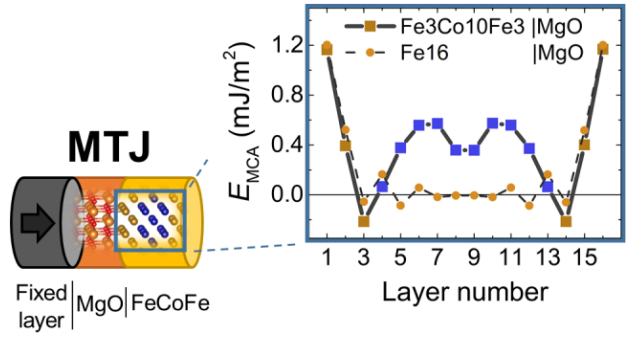

*For Table of Contents Only*

# Supporting Information

## Giant perpendicular magnetic anisotropy enhancement in MgO-based magnetic tunnel junction by using Co/Fe composite layer

Libor Vojáček[a*†], Fatima Ibrahim[a], Ali Hallal[a], Bernard Dieny[a], Mairbek Chshiev[a,b*]

[a]Univ. Grenoble Alpes, CNRS, CEA, Spintec, 38000 Grenoble, France
[b]Institut Universitaire de France (IUF)

*E-mail: Libor.Vojacek@vutbr.cz, mair.chshiev@cea.fr
†Present Address: CEITEC BUT, Brno University of Technology, Purkyňova 123, 612 00 Brno, Czech Republic

## DFT calculation method

The density functional theory (DFT) calculations were performed using the Vienna ab initio simulation package (VASP)[1,2] with generalized gradient approximation[3] and Perdew-Burke-Ernzerhof (PBE) pseudopotentials[4]. We are following the procedure described by Hallal et al.[5] with a k-point mesh of at least 25 x 25 x 1 points and a plane wave cutoff energy 520 eV. The calculation consists of three steps. First, the cell shape, volume, and atomic positions are relaxed until the forces are below 0.001 eV/Å. Second, a self-consistent calculation of the electronic structure is performed. Finally, the spin-orbit interaction is included and the energy of the system evaluated non-self-consistently with the electronic charge density from the previous step and the magnetization pointing in-plane and out-of-plane. The magnetocrystalline energy $E_{MCA}$ is then defined as the energy difference between this in-plane and out-of-plane case.

# Dipole-dipole energy calculation

The demagnetizing dipole-dipole energy $E_{dd}$ is obtained as a difference of the sum of all dipole-dipole interactions when the magnetic moments are pointing in-plane and when they are pointing out-of-plane[6,7]. We proceed as follows: First, the atomic positions of the relaxed unit cell and the magnetic dipole moments of individual atoms are loaded from the DFT calculation. Sum over all the dipole-dipole interactions is calculated up to cut-off radius $r$ (~100 unit cells), with 2D periodic boundary conditions applied within the plane of the thin film. Factor of ½ is included to avoid double counting. The difference of the magnetization in-plane and out-of-plane cases gives the $r$ dependence of $E_{dd}$. We do this for several radii $r$ and fit the resulting function $E_{dd}(r)$ with $ar^{-b} + c$. Finally, we take the $r \to \infty$ limit, i.e. $E_{dd}(r \to \infty) = c$. Note that $b \approx 1$, since the dipole-dipole energy $\propto 1/r^3$ and the number of atoms $\propto r^2$, thus overall yielding $E_{dd} \propto 1/r$.

# The minimal thickness requirement

In the Fe(n)Co(m)Fe(n)|MgO structure, at least two layers of Fe at the interface with MgO and at least three layers of Co in the bulk are needed to ensure the large perpendicular magnetocrystalline anisotropy (PMA). This is clear from Figure S1, where the layer-resolved contribution to $E_{MCA}$ is shown for various metal|MgO structures with 7 monolayers (ML) of metal.

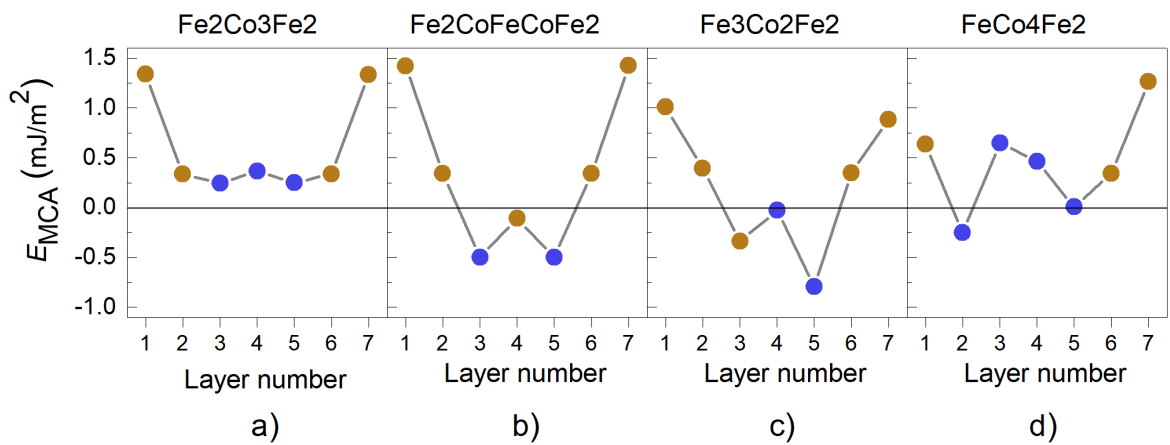

**Figure S1:** Layer-resolved $E_{MCA}$ for 7ML structures with different composition of Fe (gold) and Co (blue) layers. Comparing with (a), there is a significant drop in the $E_{MCA}$ contributions when the n=2, m=3 requirement is not fulfilled.

# PMA for structures with MgO or vacuum at the interface

In Fig. S2 we show the impact of replacing the MgO on one or both sides of the metal with vacuum and subsequent atomic relaxation on the resulting c/a and $E_{MCA}$. For the small structure with 7ML of metal, replacing the MgO with vacuum even on one of the sides is fatal for the PMA. The thicker structure is robust against this change.

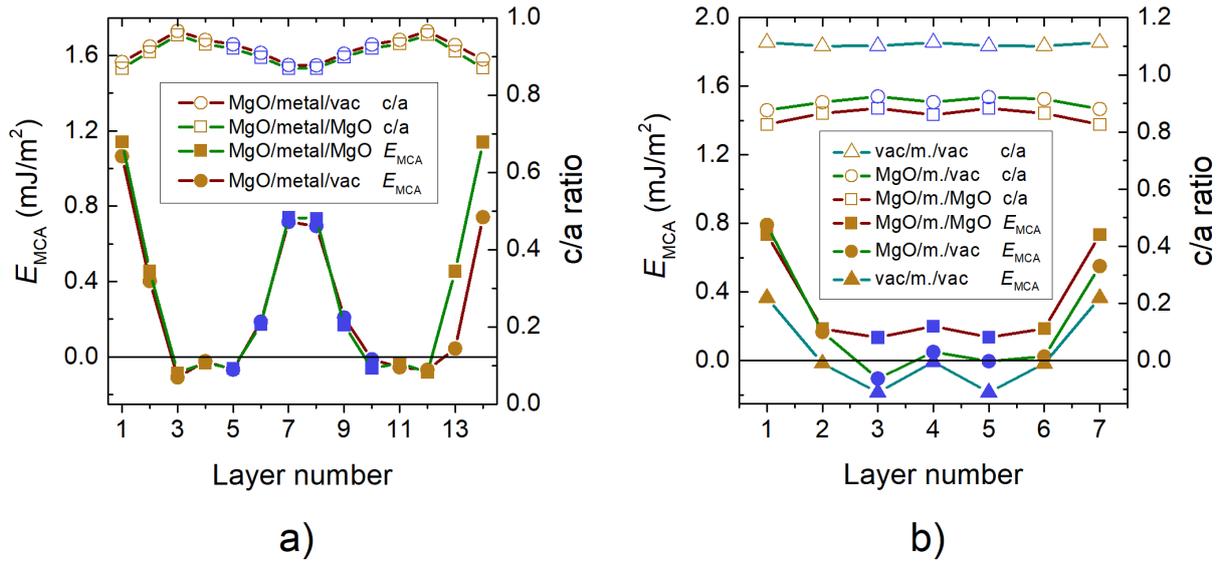

**Figure S2:** Layer-resolved $E_{MCA}$ and c/a in (a) Fe4Co6Fe4 and (b) Fe2Co3Fe2. Replacing MgO on one side with vacuum is not a big problem for the thicker structure in (a) as it decreases only the contribution from the two interfacial Fe atoms (full squares vs. full circles). However, when the thicknesses are minimal (i.e. 2ML for Fe and 3ML for Co) as in (b), the positive bulk Co contribution disappears (full squares vs. full circles). From the (layer-resolved) c/a ratio it is clear that with vacuum on one side the structure relaxes so that c/a is closer to 1 [empty squares vs. empty circles in both (a) and (b)], which in turn should decrease the PMA [Fig. 2(c)]. However, the observed decrease in (b) is radically larger than what would be expected from Fig. 2(c), suggesting an additional influence of the vacuum interface on the bulk for thinner structures. After replacing the MgO with vacuum on both sides, the structure relaxes toward c/a > 1 [empty triangles in (b)], gaining negative $E_{MCA}$ in agreement with Fig. 2(c).

# Difference in MgO and vacuum interface in bcc Fe(Co,Ni)|MgO

The effect of replacing the MgO at the interface with vacuum is shown in Figure S3. The structure was taken as relaxed with the MgO. Relaxation directly in vacuum was performed for the Fe (denoted as 'vacuum relaxed'). When performing the relaxation in vacuum for Co and Ni, the c lattice parameter expands by a factor of $\sqrt{2}$, effectively giving an fcc structure (rotated by 45° along the c axis).

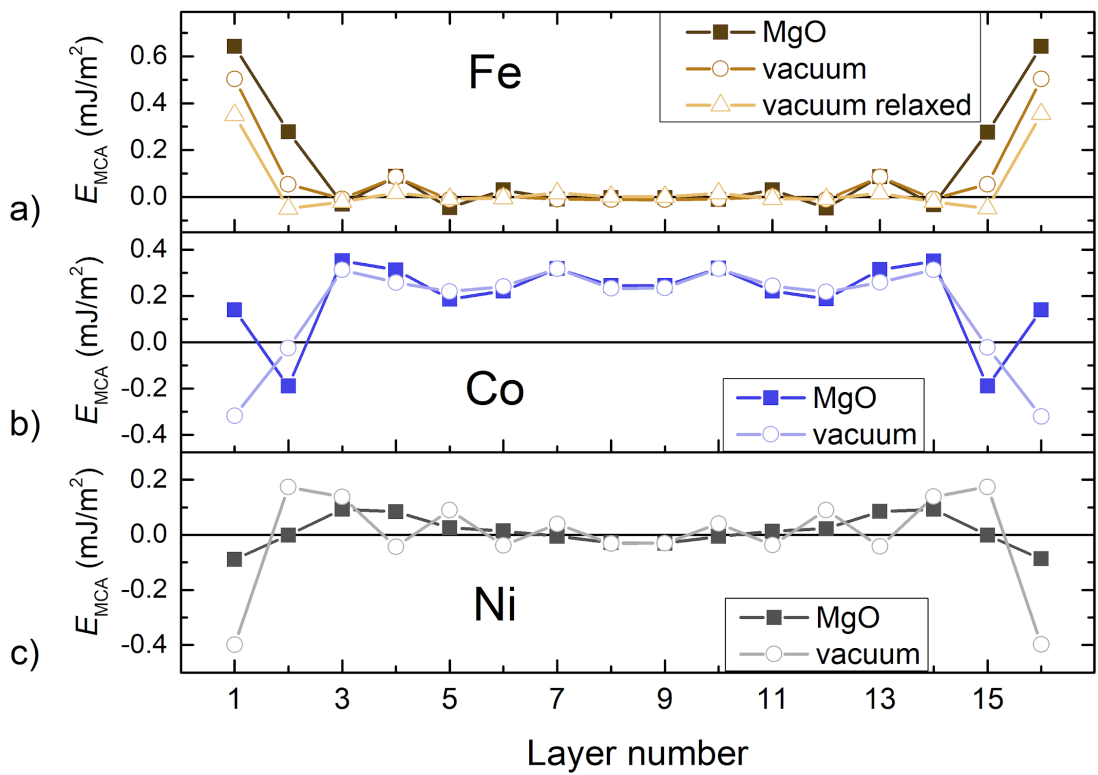

**Figure S3:** The layer-resolved $E_{MCA}$ for bcc (a) Fe, (b) Co, and (c) Ni with MgO or vacuum at the interface keeping the structure as relaxed with MgO. For Fe, the results after relaxation in vacuum are also shown. Notice the drop in the $E_{MCA}$ of the two interfacial Fe layers.

# Layer-resolved $E_{MCA}$ in enhanced structures

In Figure S4, the layer-resolved $E_{MCA}$ is shown for structures with various n and m. It is clear that adding more Fe layers decreases the PMA. On the other hand, it provides robustness against the Fe|Co interface not being atomically sharp.

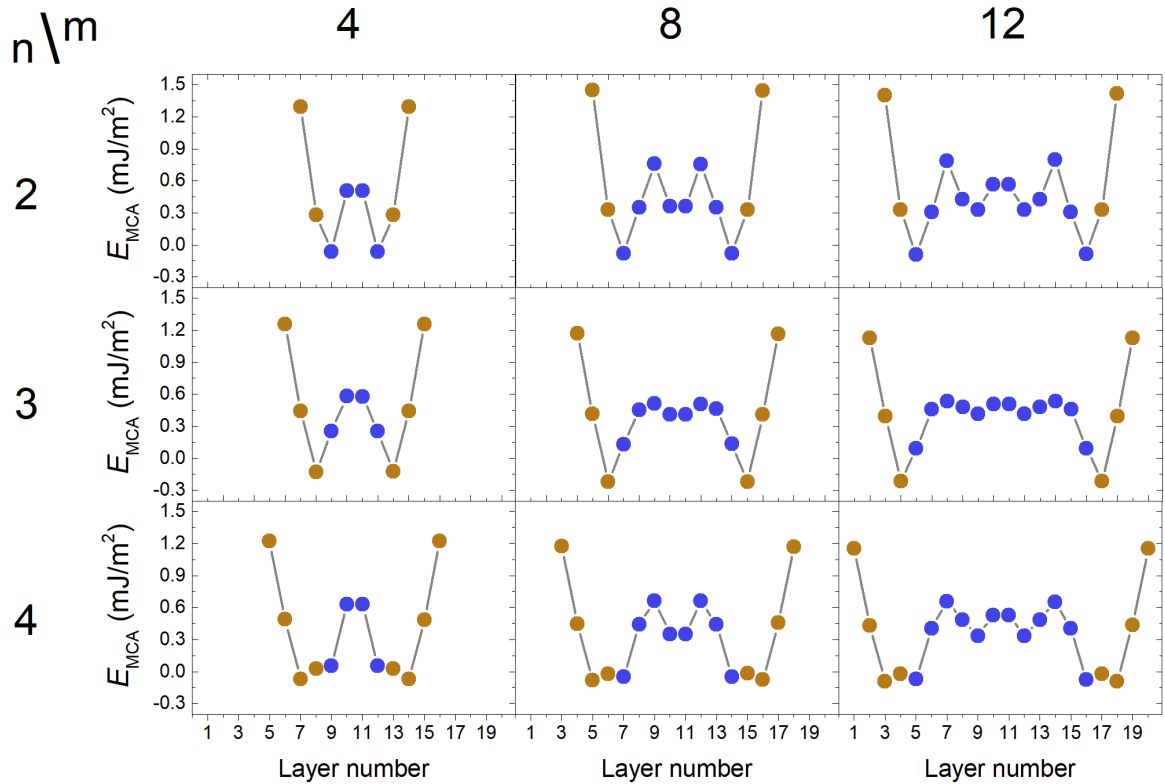

**Figure S4:** Layer-resolved $E_{MCA}$ for MgO|Fe(n)Co(m)Fe(n)|MgO structures with different n and m. The third and deeper Fe layers have almost zero contribution and decrease also the $E_{MCA}$ of the neighboring Co layer.

# Layer-resolved $E_{MCA}$ in bcc Fe(Co,Ni)|MgO of different thicknesses

In Fig. S5, one can see that while for bcc Fe|MgO and bcc Co|MgO the interfacial contributions to $E_{MCA}$ do not change significantly with thickness, in bcc Ni|MgO this is not the case.

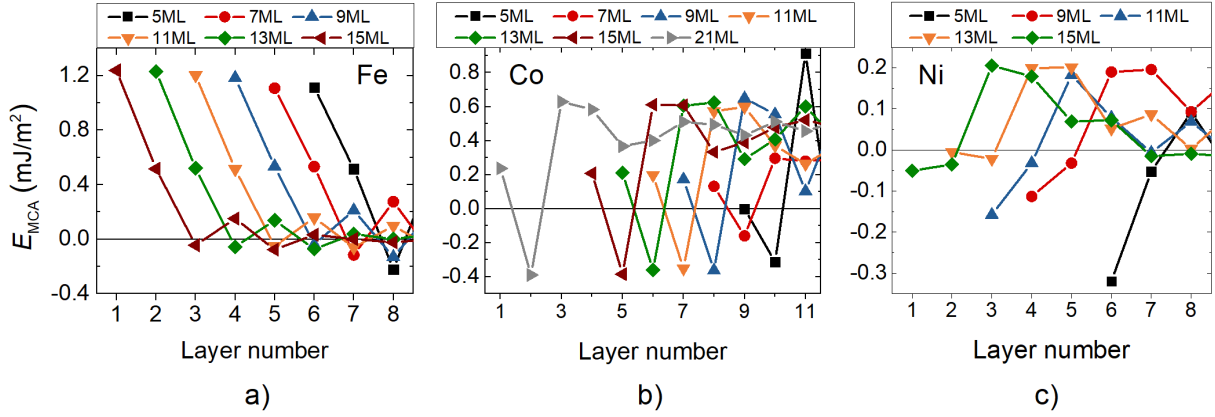

**Figure S5:** Layer-resolved $E_{MCA}$ for bcc (a) Fe|MgO, (b) Co|MgO, and (c) Ni|MgO of different thicknesses. The non-monotonic dependence of $E_{MCA}$ on thickness for the case of Ni from Fig. 2(a) can be explained by the large changes in the Ni interfacial contributions.

# The $P_{\mu\mu'}$ matrix for Bruno's model calculations

In Table 1 we show the $P_{\mu\mu'}$ matrix used in our calculations according to Bruno's model[1,2] to obtain the magnetocrystalline anisotropy energy directly from the band structure or density of states. It is obtained as $P_{\mu\mu'} = |\langle\mu|L_z|\mu'\rangle|^2 - |\langle\mu|L_x|\mu'\rangle|^2$, where $\mu$ ($\mu'$) indicates an occupied (unoccupied) $d$ orbital and $L_x$ ($L_z$) represents the orbital momentum operator in the x (y) direction. The expectation values $\langle\mu|L_i|\mu'\rangle$ can be found in Ref. 3.

| $P_{\mu\mu'}$ | d$_{xy}$ | d$_{yz}$ | d$_{z2}$ | d$_{xz}$ | d$_{x2}$ |
|---|---|---|---|---|---|
| d$_{xy}$ | 0 | 0 | 0 | -1 | 4 |
| d$_{yz}$ | 0 | 0 | -3 | 1 | -1 |
| d$_{z2}$ | 0 | -3 | 0 | 0 | 0 |
| d$_{xz}$ | -1 | 1 | 0 | 0 | 0 |
| d$_{x2}$ | 4 | -1 | 0 | 0 | 0 |

**Table 1:** The $P_{\mu\mu'}$ matrix for the $d$ orbitals. The interaction of occupied/unoccupied d$_{yz}$ and d$_{z2}$ orbitals has the largest negative contribution and it plays the central role in the PMA of strained Co.

## Supporting Information References:

(1) Kresse, G.; Hafner, J. *Ab Initio* Molecular Dynamics for Liquid Metals. *Phys. Rev. B* **1993**, *47* (1), 558–561. https://doi.org/10.1103/PhysRevB.47.558.

(2) Kresse, G.; Furthmüller, J. Efficiency of Ab-Initio Total Energy Calculations for Metals and Semiconductors Using a Plane-Wave Basis Set. *Comput. Mater. Sci.* **1996**, *6* (1), 15–50. https://doi.org/10.1016/0927-0256(96)00008-0.

(3) Wang, Y.; Perdew, J. P. Correlation Hole of the Spin-Polarized Electron Gas, with Exact Small-Wave-Vector and High-Density Scaling. *Phys. Rev. B* **1991**, *44* (24), 13298–13307. https://doi.org/10.1103/PhysRevB.44.13298.

(4) Perdew, J. P.; Burke, K.; Ernzerhof, M. Generalized Gradient Approximation Made Simple. *Phys. Rev. Lett.* **1996**, *77* (18), 3865–3868. https://doi.org/10.1103/PhysRevLett.77.3865.

(5) Hallal, A.; Dieny, B.; Chshiev, M. Impurity-Induced Enhancement of Perpendicular Magnetic Anisotropy in Fe/MgO Tunnel Junctions. *Phys. Rev. B* **2014**, *90* (6), 064422. https://doi.org/10.1103/PhysRevB.90.064422.

(6) Draaisma, H. J. G.; de Jonge, W. J. M. Surface and Volume Anisotropy from Dipole-dipole Interactions in Ultrathin Ferromagnetic Films. *J. Appl. Phys.* **1988**, *64* (7), 3610–3613. https://doi.org/10.1063/1.341397.

(7) Daalderop, G. H. O.; Kelly, P. J.; Schuurmans, M. F. H. First-Principles Calculation of the Magnetocrystalline Anisotropy Energy of Iron, Cobalt, and Nickel. *Phys. Rev. B* **1990**, *41* (17), 11919–11937. https://doi.org/10.1103/PhysRevB.41.11919.